\documentclass[twocolumn,aps,reprint,showpacs]{revtex4}

\usepackage{graphicx} 
\usepackage{dcolumn}
\usepackage{bm}
\usepackage{color}
\usepackage{amsmath}
\usepackage{amssymb}
\usepackage{textcomp}
\usepackage{natbib}
\usepackage{gensymb}

\begin{document}

\title{Decoupling thermoelectric coefficients of multilayer graphene by nanomeshing}

\author{M. Rahimi}
\affiliation{Université Paris Cité, CNRS, Laboratoire Matériaux et Phénomènes Quantiques, F-75013, Paris, France}
\author{N. Lubertino}
\affiliation{Université Paris Cité, CNRS, Laboratoire Matériaux et Phénomènes Quantiques, F-75013, Paris, France}
\author{R. Bellelli}
\affiliation{Université Paris Cité, CNRS, Laboratoire Matériaux et Phénomènes Quantiques, F-75013, Paris, France}
\author{L. Chen}
\affiliation{Université Paris Cité, CNRS, Laboratoire Matériaux et Phénomènes Quantiques, F-75013, Paris, France}
\author{F. Mallet}
\affiliation{Université Paris Cité, CNRS, Laboratoire Matériaux et Phénomènes Quantiques, F-75013, Paris, France}
\affiliation{Sorbonne Université, CNRS, UFR925, F-75013 Paris, France}
\author{P. Lafarge}
\affiliation{Université Paris Cité, CNRS, Laboratoire Matériaux et Phénomènes Quantiques, F-75013, Paris, France}
\author{C. Barraud}
\affiliation{Université Paris Cité, CNRS, Laboratoire Matériaux et Phénomènes Quantiques, F-75013, Paris, France}
\author{P. Martin}
\affiliation{Université Paris Cité, CNRS, Laboratoire ITODYS, F-75013, Paris, France}
\author{J. Chaste}
\affiliation{Centre de Nanosciences et de Nanotechnologies, Université Paris-Saclay, CNRS, 91120, Palaiseau, France}
\author{D. Fournier}
\affiliation{INSP, Sorbonne Université, CNRS, 75005 Paris, France}
\author{M. L. Della Rocca}
\email{maria-luisa.della-rocca@u-paris.fr}
\affiliation{Université Paris Cité, CNRS, Laboratoire Matériaux et Phénomènes Quantiques, F-75013, Paris, France}

\date{\today}

\begin{abstract}
Nanostructuring materials at small scales enables control over their physical properties, revealing behaviors not observed at larger dimensions.
This strategy is particularly effective in two-dimensional materials, where surface effects dominate, and is especially promising for energy conversion and thermal management.
Here, we use multilayer graphene (4-6 nm thick) as a test platform to study the effect of nanomeshing on its thermoelectric coefficients. The nanomesh consists of a hexagonal array of holes, with a measured diameter and neck-width of $\sim$ 360 nm and $\sim$ 160 nm, respectively. The multilayer graphene is integrated into field-effect transistor-like devices supported by hexagonal boron nitride, allowing simultaneous electric and thermopower measurements, with nanomeshing applied to only part of the material. We use modulated thermoreflectance to investigate thermal transport in equivalent nanomeshed and pristine graphene flakes. Our results show a reduction in electrical conductivity that can be well described within the framework of the classical Maxwell–Eucken model for porous media. In contrast, the observed reduction in thermal conductivity exceeds what expected from geometrical modifications alone, highlighting the different mean free paths of phonons and electrons. Moreover, we find an increased thermopower in the nanomeshed graphene region, which can be qualitatively attributed to filtering of low-energy charge carriers.
These findings demonstrate that nanomeshing can significantly decouple the thermoelectric coefficients of multilayer graphene, paving the way for novel engineering strategies to achieve similar control across a wider range of 2D materials.
\end{abstract}


\maketitle 
\section{\label{sec:one} Introduction}
In the past decades, innovative concepts have emerged, suggesting new strategies for engineering and improving material properties through reduced dimensionality \cite{hicks_Lyndon_D_1993_effect, dresselhaus_Mildred_S_2007_new, dresselhaus_DS_Dresselhaus_1999_low}. Low-dimensional systems, including van der Waals (vdW) two-dimensional (2D) materials, exhibit unique electron and phonon densities of states (DOS), which can lead to enhanced thermoelectric performance \cite{dehkordi_Arash_Mehdizadeh_2015_thermoelectric}. Among 2D materials, graphene exhibits an extraordinary high electrical conductivity, $\sigma$, and a relatively large Seebeck coefficient, $S$, measured up to $\sim$ 200 µV/K \cite{duan_Junxi_Wang_2016_high, zuev_Yuri_M_2009_thermoelectric}, due to the energy dependence of the DOS near the charge neutrality point \cite{novoselov_Kostya_s_2005_two}, allowing for high power factors, $PF=S^2 \sigma$. 
However, the use of graphene as a thermoelectric material is limited by its extremely high thermal conductivity which reduces the energy conversion efficiency.
The strong in-plane covalent bonds allow phonon mean free paths of up to several hundred nanometers \cite{ghosh_Suchismita_Calizo_2008_extremely, xu_Xiangfan_Pereira_2014_length}, resulting in an in-plane thermal conductivity, $k$, higher than 3000 W/m K \cite{Cai2010, ghosh_Suchismita_Calizo_2008_extremely} at room temperature. This makes graphene more suitable as an efficient heat spreader. 
Due to the different mean free paths of electrons and phonons in graphene, recent research has focused on developing techniques aimed at decoupling the thermoelectrics coefficients $k$, $S$, and $\sigma$.  \cite{nobakht_Ali_Yousefzadi_2017_heat}. Among the various approaches for tuning these properties—such as edge passivation \cite{hu_Jiuning_Schiffli_2010_tuning}, introduction of defects \cite{haskins_Justin_Kinaci_2011_control}, grain boundary engineering \cite{serov_Andrey_Y_2013_effect} and hybridization methods \cite{Kim2025}—creating periodic hole structures (graphene nanomesh, GNNM) stands out as one of the most practical methods \cite{oh_jinwoo_yoo_2017_significantly, Li2019}. Nanostructuring \cite{feng_Tianli_Ruan_2016_ultra} and substrate engineering \cite{duan_Junxi_Wang_2016_high} have been demonstrated as efficient methods, for controlling thermoelectric parameters \cite{oh_jinwoo_yoo_2017_significantly}. However, these approaches can also significantly reduce electrical conductivity \cite{bai_Jingwei_Zhong_2010_graphene}, making careful tuning of parameters essential.
For instance, it has been shown experimentally, that nanomesh patterns in graphene at sub-20 nm scales significantly alter its charge transport properties due to quantum confinement and the formation of electronic sub-bands \cite{liang_Xiaogan_Jung_2010_formation}. Nanomeshed graphene has been mainly used as a strategy to transform it from a semimetal into a semiconductor \cite{hung_Nguyen_V_2013_disorder}, with an electronic bandgap that can be adjusted by varying the size, shape, and symmetry of both the holes and the lattice cells \cite{ouyang_Fangping_Peng_2011_bandgap}. 
On the other hand, nanostructuring is also a powerful approach to reduce thermal conductivity \cite{bai_Jingwei_Zhong_2010_graphene, ouyang_Fangping_Peng_2011_bandgap, liang_Xiaogan_Jung_2010_formation}. Theoretical calculations reveal that ballistic, size-dependent thermal transport can occur in GNNMs, leading to a significant decrease in the overall thermal conductivity as porosity increases \cite{nobakht_Ali_Yousefzadi_2017_heat}.
However, fewer experimental studies have investigated the thermal properties of GNNMs \cite{oh_jinwoo_yoo_2017_significantly, xiong_Ze_Wang_2018_thermal}, primarily using suspended structures and techniques such as the thermal bridge method \cite{xiong_Ze_Wang_2018_thermal} and optothermal Raman spectroscopy \cite{oh_jinwoo_yoo_2017_significantly}, with neck widths ranging from the sub-10 nm regime to several hundred nanometers and periodicities spanning from tens to hundreds of nanometers \cite{xiong_Ze_Wang_2018_thermal}. In all cases, the reduction of the thermal conductivity can be of more than a factor 10.
On the opposite, the effect of nanomeshing on graphene's thermopower has been scarcely investigated, although modifications to the band structure from nanostructuring are expected to have an impact. It has been theoretically demonstrated that the DOS modification induced by atomic-scale holes in graphene can significantly enhance the Seebeck coefficient, reaching values as high as 0.5 mV/K \cite{karamitaheri_Hossein_Pourfath_2011_geometrical}. The presence of pores is also expected to increase the Seebeck coefficient through effective filtering of low-energy carriers via potential barriers formed at the pores edges \cite{hao_Qing_Zhao_2017_thermoelectric}. Trapped charges at the pore edges in GNNM systems create a local electric field that induces an energy barrier, primarily affecting the transport of low-energy carriers. Since the Seebeck coefficient is proportional to the average energy of charge carriers relative to the Fermi level at a given temperature, this filtering effect is expected to increase the thermopower. However, due to the overall reduction in free carrier density caused by the filtering, large energy barriers can lead to a decrease in electrical conductivity too \cite{hao_Qing_Zhao_2017_thermoelectric}.

In this work, we investigate the influence of nanostructuring on the thermoelectric coefficients ($\sigma$, $S$, and $k$) of multilayer graphene. Due to its well-established electric and thermal properties and ease of manipulation, here graphene serves as an ideal 2D test-bed for exploring geometric modifications aimed at controlling the thermoelectric parameters.
We apply a nanomesh patterned consisting of a hexagonal array of holes, with pore diameters $d$ and neck width $w$ in the few hundreds of nanometers range, with the objective of minimally impacting charge transport while mostly affecting thermal transport. The FET-like device architecture enables a comprehensive comparaison of the thermoelectric coefficients of the same material, where nanomeshing is applied to only a selected region. 
We measure the thermal conductivity by modulated thermoreflectance (MTR) on equivalent supported nanomeshed and pristine graphene flakes. 
The applied nanomesh results in enhanced thermopower and suppressed effective electrical and thermal conductivities. While the reduction in the electrical conductivity is in good agreement with classical expectations for porous media, we observe a threefold decrease in the thermal conductivity, indicating a stronger impact of the chosen geometry on the phonon mean free path, going beyond the classical predictions.
These findings demonstrate that nanostructuring by a network of holes provides an effective route to significantly control the thermoelectric coefficients of multilayer graphene, opening new avenues for engineering advanced nanomeshed approaches for energy conversion technologies based on a larger panel of 2D materials.

\begin{figure*}
\includegraphics[scale=0.62]{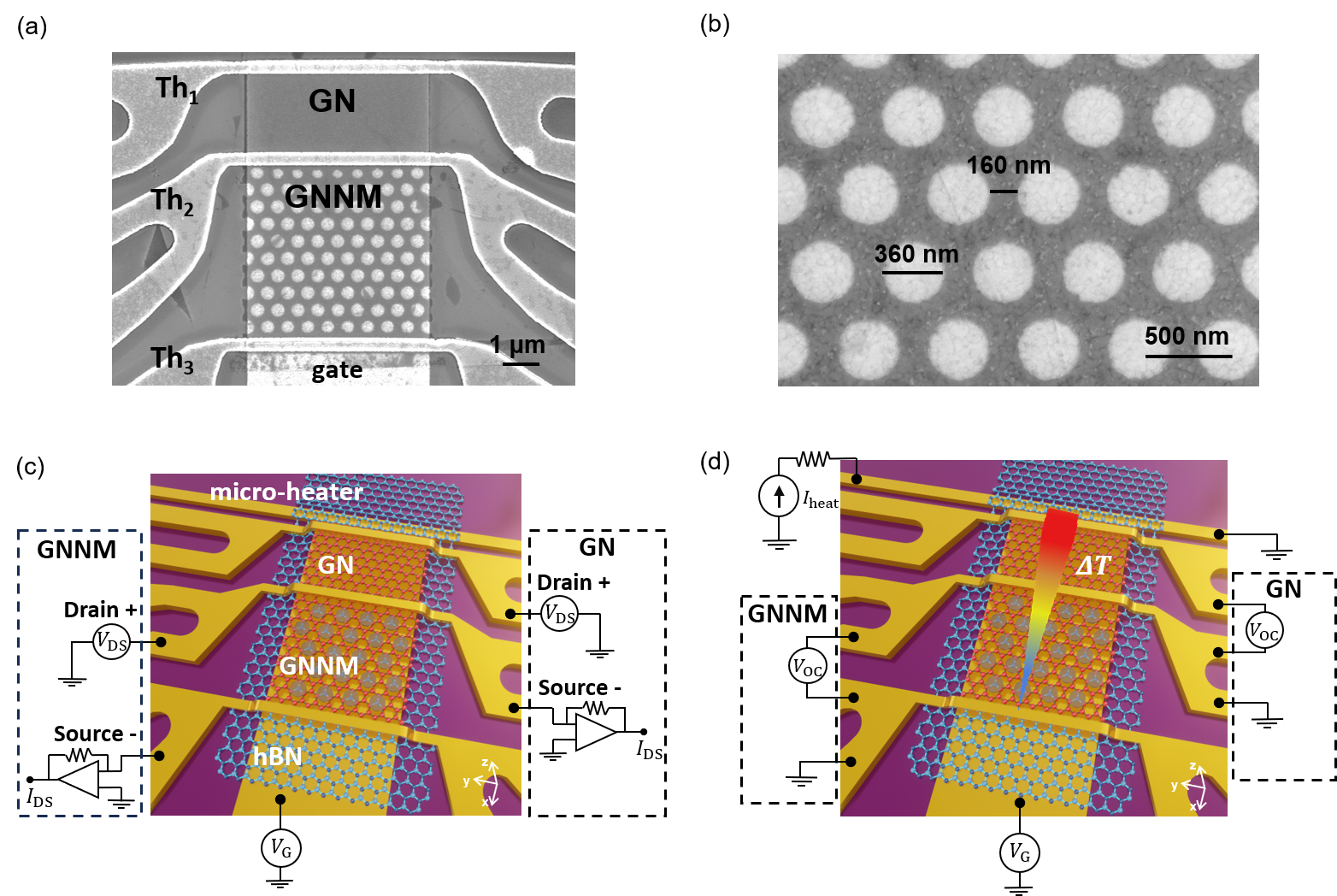}
\caption{\label{Fig1} (a) Scanning electron microcope (SEM) image of a typical FET-like device including multilayer graphene (GN) where nanomeshing has been applied to a portion of it (GNNM). (b) Zoomed-in SEM image of the nanomeshed graphene (GNNM). (c) Artistic view of the device illustrating details for the electrical transport measurements of the GN and GNNM regions. (d) Artistic view of the device illustrating details for the thermopower measurements of the GN and GNNM regions.}
\end{figure*}

\section{\label{sec:two} Devices fabrication and measurements}
Devices are fabricated using standard micro- and nano-fabrication tools \cite{rahimi_M_Sobnath_2023_complete, timpa_Salvatore_Rahimi_2021_role} on Si/SiO$_2$ (300 nm) substrates. The multilayer graphene, including a pristine (GN) and a nanomeshed (GNNM) part, is capacitevely coupled to a local metallic gate through a flake of hexagonal boron nitride (hBN) used as a dielectric spacer. 
The 2D materials are mechanically exfoliated from the bulk (HQGraphene) and subsequently dry-transferred over the metallic gate. Three metallic nanowires deposited on the top
serve as local electrodes and thermometers. The nanowires have a width of 350 nm and are made of $5~$nm Ti and $75~$nm of Au. 
The nanomeshing of a part of the multilayer graphene is performed as the final fabrication step. A thin ($\sim8~$nm) Al metallic mask is deposited by e-beam evaporation over the device and patterned into the desired nanomeshed structure. Plasma oxygen is used to etch graphene through the Al mask, which is subsequently chemically removed. During the etching process, the GN and GNNM parts are also reshaped to match the gate dimensions. 
The channel width is $5~\mu$m, while the length is $2~\mu$m for the GN part and $ 4.8~\mu$m for the GNNM part.
This device configuration allows for a complete electric and thermoelectric characterization of the GN and GNNM regions separately.
The dimensions of the pore diameters and neck width have been optimized to $d \sim$ 360 nm and $w \sim$ 160 nm, respectively, in order to preserve the electronic properties of the nanomeshed multilayer graphene as much as possible. For $w$ of the order of $\sim$100 nm and $\sim$50 nm, we observe a suppression of the effective electrical conductivity of the nanomeshed graphene of approximately one and two order of magnitude, respectively.
Note that the hBN flake under the graphene can also be partially etched after the complete removal of the unprotected graphene areas.
However, hBN exhibits a substantially slower etching rate in comparison to graphene 
\cite{hug_Dorothee_Zihlmann_2017_anisotropic, sun_Haibin_Dong_2021_etching}, as a consequence it is only slightly affected by the etching process, maintaining both its structural integrity and properties.

Here we present results on a representative device, with graphene thickness of $\sim$ 4 nm, whose scanning electron microscope (SEM) image is shown in Fig.~\ref{Fig1} (a). Analogous results have been obtained over other 3 equivalent devices, with graphene thickness ranging from $\sim$ 4 nm to $\sim$ 6 nm.  Figure~\ref{Fig1} (a) clearly shows the three electrodes (Th$_1$, Th$_2$, Th$_3$) separating the GN and GNNM regions. Finite element simulations (Comsol Multiphysics, Fig. S1 in Supplementary Information) show that the temperature gradient decays exponentially along the longitudinal direction of the device. For this reason, in order to establish a reliable temperature gradient through the GNNM part, the Th$_1$–Th$_2$ distance is set shorter than the Th$_2$–Th$_3$ distance. A zoom-in image of the GNNM region is provided in Fig.~\ref{Fig1} (b), showing the regular periodic hole structure with holes organized in a hexagonal arrangement.

A 3D artistic view of the typical FET device based on hBN/GN and GNNM is shown in Fig.~\ref{Fig1} (c) and (d), illustrating the circuit for the electric and thermoelectric measurements, respectively. The measurements are performed at 35°C under high vacuum (P $\sim$ 10$^{-7}$ mbar) in a multiprobe station (Nextron) after in-situ annealing (2 h at 300°C). This annealing step is fundamental to remove the absorbates
from the graphene surfaces after microfabrication and in particular to improve the quality of the 2D material/electrodes interfaces and the stability of the nanowires resistances. The GN and GNNM regions are fully characterized electrically and thermoelectrically, one after the other. Current-voltage (I-V) characteristics are acquired in a two-point configuration by voltage-biasing the samples (Yokogawa 7651) and measuring the induced current as a function of the gate voltage using a low-noise current-voltage amplifier (Femto DLPCA-200). Leakage currents between gate, electrodes, and micro-heater are lower than the sensitivity of our experimental setup ($\sim$ 50 pA) over the explored gate and source-drain voltage ranges.

The thermoelectric characterization is performed by inducing a temperature gradient, $\Delta$T, along the longitudinal direction (x-direction in Fig.~\ref{Fig1} (d)) by Joule heating of a local micro-heater, which is electrically isolated from the FET structure.  The temperature gradient is measured using the electrodes as local thermometers, after precisely calibrating their resistances as a function of temperature. The thermoelectric signal is obtained by measuring the open-circuit voltage, V$_{oc}$, between two electrodes using a nanovoltmeter (Keithley 2182A) as a function of the current I$_{heater}$ in the micro-heater and the gate voltage V$_G$ in the typical maximum range of $-$15 V $\leq$ V$_G$ $\leq$ 15 V.
The pure thermoelectric contribution is extracted by considering the second-order term of V$_{oc}$ in I$_{heater}$, V$_{th}$ $\propto$ I$_{heater}^2$, which is typically the dominant one. Thus, the Seebeck coefficient is obtained from the slope of the V$_{th}$ vs. $\Delta$T dependence. More details about the measurement procedure have been given elsewhere and are presented in the Supplementary Information \cite{rahimi_M_Sobnath_2023_complete, rahimi_M_Sobnath_2024_probing}.

We use modulated thermoreflectance (MTR) to investigate thermal transport properties of GN and GNNM flakes. We have recently demonstrated that MTR is a reliable nondestructive
optical pump-probe technique allowing the measurement of the in-plane and out-of-plane thermal conductivities of the different 2D flakes with thicknesses in the few nanometers range \cite{rahimi_M_Sobnath_2023_complete}.
In MTR experiments, the measured signal is proportional to the optical reflectance variation of a probe beam (blue laser, 488 nm, LBX-488 Oxxyus at $\sim$ 100 $\mu$W) as a function of temperature in the heated area created by a pump beam (green laser, 532 nm, Cobolt Samba at $\sim$ 1 mW), modulated at a frequency $f$ in the hundreds of kHz range. We measure the amplitude and phase of the probe signal by homodyne detection as a function of the distance from the pump-laser spot, following linear scans.
Data are fitted by solving a three-dimensional theoretical model of classical heat transport describing the temperature distribution in multilayered samples, heated by an intensity-modulated Gaussian laser beam, including both lateral and vertical heat diffusion \cite{fournier_Daniele_Marangolo_2020_measurement, li_Bin_cheng_1997_effect, li_Bincheng_Roger_1999_complete}.
More details about the measurements and data analysis can be found in Refs. \cite{fournier_Daniele_Marangolo_2020_measurement, rahimi_M_Sobnath_2023_complete}.
Given the reduced dimensions of the devices, MTR measurements are performed on separate samples with the same Au gate/hBN/GN and GNNM flakes staking, embedding 2D materials with thicknesses similar to the ones used for the electric and thermoelectric characterization, but with longer channel lenght. 

\begin{figure*}
\includegraphics[scale=0.55]{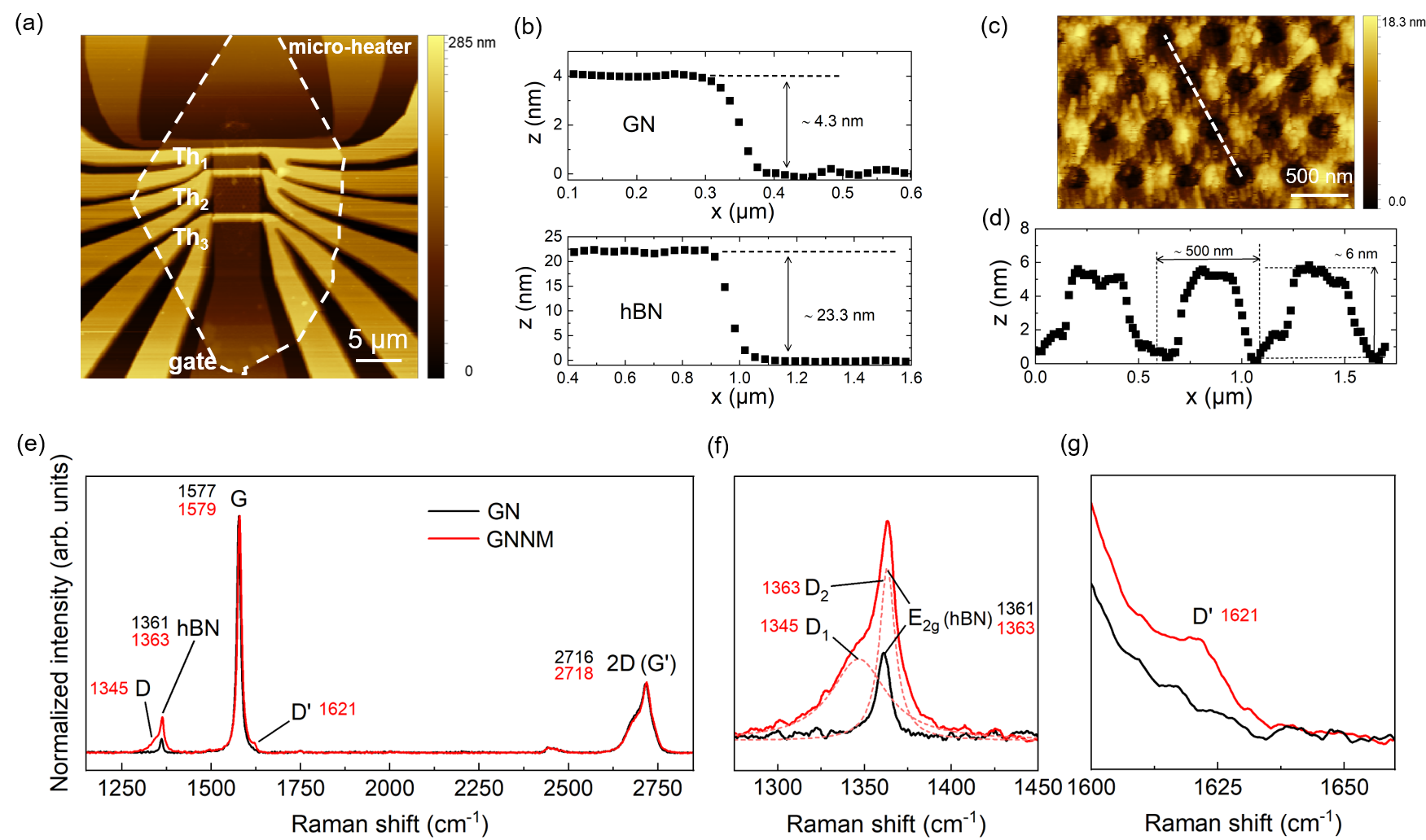}
\caption{\label{Fig2} (a) Atomic force microscopy (AFM) image of the presented device indicating the different parts (electrodes, micro-heater, gate, flakes). The border of the hBN flake is highlithed by the dashed white  line. (b) Averaged AFM line profile of the hBN flake (bottom panel) and GN flake (top panel) of the device in (a) allowing the flakes thicknesses extraction. (c) Zoomed-in AFM image of the GNNM region. (d) AFM line profile cut along the with dotted line in (c) allowing to extract a hole depth of $\sim$ 6 nm. (e) Raman spectra ($\lambda$=532 nm, Witec) of GN (black data) and GNNM (red data) parts respectively, showing multiple well-defined peaks. The correspondent phonon modes are indicated. (f) Zoom of the Raman spectra around the D band at 1362 cm$^{-1}$ allowing to detect the signature of an increased disorder in the GNNM. (g) Zoom of the Raman spectra around the D' band at 1621 cm$^{-1}$ visible only in the spectra of GNNM (red curve) confirming an increased disorder. Note that not always the GN and GNNM spectra are distinghuishable, the presented data are related to a different sample than the one discussed.}
\end{figure*}

\section{\label{sec:two} Flake structural characterization}
Prior to the nanostructuring process, atomic force microscopy (AFM) is performed to determine the thickness and surface roughness of the multilayer graphene. A second AFM analysis is carried out after nanostructuring to confirm the complete etching of GN within the holes.
Figure~\ref{Fig2} (a) shows the AFM image of the presented device after nanomeshing, where the boundary of the hBN flake is indicated by a white dotted line. We clearly distinguish the different device parts (electrodes, micro-heater, gate). Flake thicknesses are investigated on the AFM images acquired after tranfer of hBN and graphene, before the electrode deposition and nanomeshing. The thicknesses are extracted by averaging the step heights from profiles taken in at least 50 different zoomed-in regions.
Figure~\ref{Fig2} (b) illustrates representative height profiles of multilayer graphene (top panel) and hBN (bottom panel) of the discussed device. The measured thickness of the multilayer graphene is 4.3 ± 0.7 nm, while the hBN flake thickness is 23.3 ± 0.9 nm. The roughness, averaged over an area of 1 $\mu$m$^2$, is approximately 1.1 nm for graphene and 0.85 nm for hBN.
Although the holes are faintly visible in Fig.~\ref{Fig2} (a), they are clearly resolved in the AFM image shown in Fig.~\ref{Fig2} (c), which corresponds to a zoomed-in scan of the GNNM region. A line profile along the white dotted line in Fig.~\ref{Fig2} (c) is plotted in Fig.~\ref{Fig2} (d) and reveals an etching depth of approximately 6 nm, confirming that the graphene is completely etched within the holes, and that few nanometers of the hBN have been removed. The center-to-center distance estimated from the AFM hole profiles closely matches the expected value of $\sim$ 500 nm. However, due to the curvature radius of the AFM tip, the accuracy of the hole diameter and neck width measurements is lower than that obtained from SEM analysis.

Raman spectroscopy is employed to investigate the properties of the GN and GNNM flakes, as it is a powerful technique for detecting disorder and doping in graphene \cite{eckmann_Axel_Felten_2013_raman, ferrari_Andrea_C_2007_raman_spectroscopy}. Figure~\ref{Fig2} (e) shows the Raman spectra of the GN flake (black curve) and the GNNM flake (red curve), measured at room temperature using a 532 nm laser excitation (WiTec) on an analogous device. 
The Raman spectrum of GNNM reveals the presence of structural disorder induced by the hole edges, causing subtle spectral changes compared to the pristine multilayer graphene. Not always differences in Raman spectra of the GN and GNNM regions are clearly observable, as it is the case for the spectra of the device in Fig.~\ref{Fig1} (a), where disordered induced peaks are present in both regions. For spectra shown in Fig.~\ref{Fig2} (e-g), these changes are most evident.
Both GN and GNNM spectra are normalized to their respective maximum intensity values.
For both the GN and GNNM flakes, two prominent peaks appear at $\sim$ 1578 and 2717 cm$^{-1}$, corresponding to the G and 2D bands of multilayer graphene \cite{ferrari_Andrea_C_2006_raman_spectrum}. 
The presence of both G and 2D bands indicates that the sp$^{2}$ carbon network remains intact, despite the patterning and etching process \cite{ferrari_Andrea_C_2007_raman_spectroscopy}.
Additionally, the higher intensity of the G band relative to the 2D band ($I_G/I_{2D}$ $>$ 1) confirms the multilayer nature of the graphene flakes \cite{ferrari_Andrea_C_2006_raman_spectrum}.
In the GNNM spectrum, a pronounced peak at 1362 cm$^{-1}$ is observed (Fig.~\ref{Fig2} (f), red curve). This peak is also present in the GN spectrum (Fig.~\ref{Fig2} (f), black curve) and is attributed to the E$_{2g}$ mode of the underlying hBN layer \cite{stenger_I_Schue_2017_low}. In the GNNM spectrum, the peak appears more intense and broader and can be decomposed into three contributions: the E$_{2g}$ mode of the hBN layer and a defect-induced D band composed of two components, D$_1$ and D$_2$, at $\sim$ 1345 cm$^{-1}$ and $\sim$ 1363 cm$^{-1}$, respectively (red dashed lines) \cite{ferrari_Andrea_C_2006_raman_spectrum, malard_Leandro_M_2009_raman}. The D band, associated with in-plane transverse optical (TO) phonon scattering, is normally Raman-inactive but becomes observable through defect-induced scattering \cite{eckmann_Axel_Felten_2012_probing}.
Therefore, the enhanced intensity of the $\sim$ 1362 cm$^{-1}$ peak in the GNNM flake is indicative of increased structural disorder, likely due to partial etching of the graphene. This conclusion is further supported by the appearance of the D' band around 1621 cm$^{-1}$, visible only in the GNNM spectrum (Fig.~\ref{Fig2} (g), red curve). This band is attributed to intravalley scattering of longitudinal optical (LO) phonons \cite{eckmann_Axel_Felten_2013_raman, eckmann_Axel_Felten_2012_probing}, typically caused by defects in the graphene crystal lattice.

\section{\label{sec:three} Results and discussion}
\begin{figure*}
\includegraphics[scale=0.53]{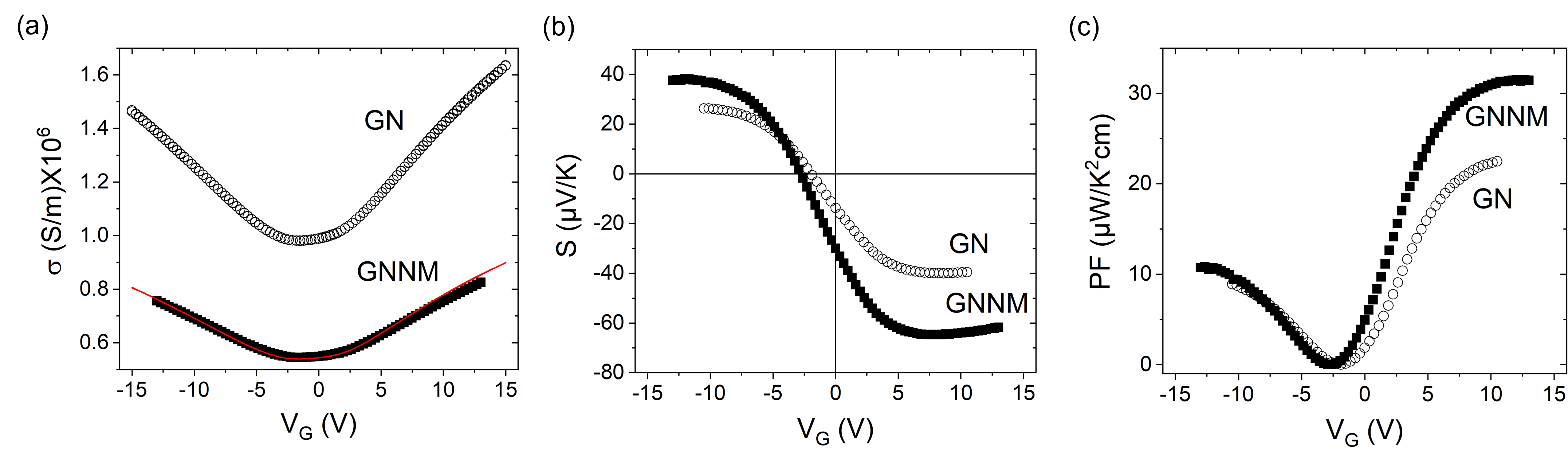}
\caption{\label{Fig3} (a) Electrical conductivity $\sigma$ of the GN (open symbols) and GNNM regions (solid symbols) as a function of the gate voltage V$_G$. The continous red line is the expected $\sigma$ redicton calculated by the Maxwell-Eucken model. (b) Seebeck coefficient $S$ for the GN region (open symbols) and the GNNM region (solid symbols) as a function of the gate voltage, showing an enhanced response in the presence of the nanomeshed network. (c) Power factor $PF=S^2\sigma$ for the GN region (open symbols) and the GNNM region (solid symbols) as a function of the gate voltage, showing enhanced values in the nanomeshed case.} 
\end{figure*}

We now discuss the electric and thermoelectric characterization of the device in Fig.~\ref{Fig1} (a). 
The current–voltage (I–V) characteristics of both GN and GNNM regions display a linear trend, as shown in Fig.~S2 (a) in the Supplementary Information. The measured resistances are approximately $\sim 89~\Omega$ for the GN region and $\sim 394~\Omega$ for GNNM region.
We additionally measure the contact resistance of the central electrode (Th$_2$) using a three-point method and obtain a value of $\sim 6~\Omega$. This result indicates quite good ohmic contacts, which do not dominate transport measurements. Since all the nanowires were fabricated simultaneously prior to nanomeshing, it is reasonable to assume that the graphene/electrode interfaces exhibit similar properties.
Transconductance measurements are conducted at a fixed $\sim 6$~mV source-drain voltage for both regions, while sweeping the gate voltage (Fig.~S2 (b) in Supplementary Information). 
We apply Ohm’s second law to extract the electrical conductivity of the GN ($\sigma_{GN}$) and GNNM ($\sigma_{GNNM}$) regions. While this approach is exact for the GN part which has a well-defined rectangular geometry, in the case of the GNNM part, $\sigma_{GNNM}$ has to be considered as an effective physical quantity. 
Figure~\ref{Fig3} (a) presents the extracted $\sigma_{GN}$ and $\sigma_{GNNM}$ as a function of the applied gate voltage $V_G$ (open and solid symbols, respectively).

In both cases the electrical conductivities are slightly modulated by the gate voltage. The electrical conductivity of the GNNM region is quite reduced with respect to the GN region. The extracted $\sigma_{GNNM}$ values vary from $0.5 \times 10^6$~S/m to $0.8 \times 10^6$~S/m, while the extracted $\sigma_{GN}$ values range from $1.0 \times 10^6$~S/m to $1.6 \times 10^6$~S/m. 
The minimum value of $\sigma$ occurs in both cases at the same slightly negative gate voltage $(\approx -2~V)$. The similar behavior observed in the gate-dependent transport properties is closely related to the fact that both the GN and GNNM regions originate from the same graphene flake.
The conductivity modulation cannot be assigned to a well-defined energy gap, but rather to the complex band structure of multilayer graphene, which is explored by varying the gate voltage. Alternatively, the conductivity modulation can be explained by a gate-induced effect confined to very few layers of the GN and GNNM regions, where charge carrier density strongly screen the gate electric field. Note that the gate voltage range is intentionally limited to values below $\sim \mid15\mid$~V to avoid leakage through the hBN layer.
We model the GNNM region as a porous medium containing vacuum pores, whose electrical and thermal properties can be described classically within the framework of the Maxwell–Eucken model \cite{oh_jinwoo_yoo_2017_significantly, tang_Jinyao_Wang_2010_holey}. This effective medium approach accounts for the impact of inclusions (here, pores) on the electric and thermal transport by treating them as non-conducting regions embedded in a homogeneous matrix. It provides analytical expressions linking the effective electrical and thermal conductivities to the porosity. In this framework, for a given porosity $P$, the estimated reduction of the electrical (and thermal) conductivity is given by $\sigma_{\mathrm{GNNM}}/\sigma_{\mathrm{GN}} = (1-P)/(1+P/2) \approx 0.55$. This estimation is in good agreement with the experimental data. In fact, by applying a factor of 0.55 to the measured gate dependence of $\sigma_{GN}$, we obtain values that closely match the measured gate dependence of $\sigma_{GNNM}$ (continuous red line in Fig. ~\ref{Fig3} (a)).

The Seebeck coefficient of the GN and GNNM regions is measured shortly after the electrical characterization at a fixed temperature T$= 35^\circ \text{C}$ as a function of the gate voltage. For each gate voltage V$_G$ applied to the local-gate electrode, a temperature gradient $\Delta$T is generated by sweeping the heating current I$_{\text{heater}}$ in the micro-heater between 0 and $\sim$ 4 mA, and is precisely measured using the 3 nanowires as local thermometers. We typically measure a maximum $\Delta$T $\sim$ 7-8 K across the GN part and $\Delta$T $\sim$ 1-2 K across the GNNM part. Details of the thermometer calibration procedure and the measurement of the temperature gradient can be found in Fig. S3 in the Supplementary Information. For each V$_G$ value, the Seebeck coefficient is given by the linear slope of V$_{\text{th}}$ as a function of $\Delta$T \cite{rahimi_M_Sobnath_2024_probing, timpa_Salvatore_Rahimi_2021_role}.
Figure~\ref{Fig3} (b) shows the Seebeck coefficient $S$ for the GN region (open symbols) and the GNNM region (solid symbols) as a function of the gate voltage, demostrating an enhanced response across the entire gate voltage range in the presence of the nanomeshed network. 
The Seebeck coefficient for both devices exhibits a sign change, consistent with the electrical conductivity behavior shown in Fig.~\ref{Fig3} (a). The sign of $S$ indicates the nature of the majority charge carriers contributing to the thermoelectric voltage. An intrinsic $n$-doping is present in both cases, as previously reported in literature \cite{xu_Dongchao_Tang_2019_detecting, timpa_Salvatore_Rahimi_2021_role}.  
Note that the measured $S$ values are comparable in magnitude to those observed for single-layer graphene \cite{zuev_Yuri_M_2009_thermoelectric, oh_jinwoo_yoo_2017_significantly}. 
Note also that the $S$ measurements is asymmetric between the electron and hole regimes, while a similar but weaker asymmetry is present in the electrical conductivity. This likely originates from subtle variations in the regions of the device contributing to each measurement: electronic transport could be dominated by the surface layers, whereas the thermoelectric voltage is mainly governed by the whole layered structure. 
The GN region achieves a maximum absolute value of the Seebeck coefficient of $\sim$ 40 $\mu$V/K in the electron transport regime and $\sim$ 27 $\mu$V/K in the hole transport regime. In contrast, for the GNNM region the Seebeck coefficient is $\sim$ 65 $\mu$V/K in the electron transport regime and 40 $\mu$V/K in the hole transport regime. Consequently, we estimate an increase of $S$ by approximately 60\% in the first case and 48\% in the second case. We measure an increase of the Seebeck coefficient in presence of pores consistent with an effective filtering of low energy carriers. 

With both $S$ and $\sigma$, we evaluate the power factor $PF = S^2 \sigma$, as a function of the gate voltage for both the GN and GNNM regions, as shown in Fig.~\ref{Fig3} (c). The GNNM part exhibits a higher power factor than the GN part. The maximum power factors are $\sim$ 32 $\mu$W/K$^{2}$ cm and $\sim$ 22 $\mu$W/K$^{2}$ cm, for the GNNM and GN regions, respectively, in the electron transport regime. Here, the maximum thermoelectric power factor is increased by more than 40$\%$ in the presence of the nanomeshed structure, revealing a potential interest of the used nanomeshed geometry for improving active cooling efficiency of multilayer graphene, where heat is transported through the electronic channel and controlled by the applied current (Peltier effect). 
Although there is no theoretical upper limit for $PF$, the trade-off between the Seebeck coefficient and electrical conductivity has historically hindered the achievement of high thermoelectric power factors. Nanostructuring with optimized geometries offers a promising and reliable route for overcoming this limitation.

\begin{figure*}
\includegraphics[scale=0.55]{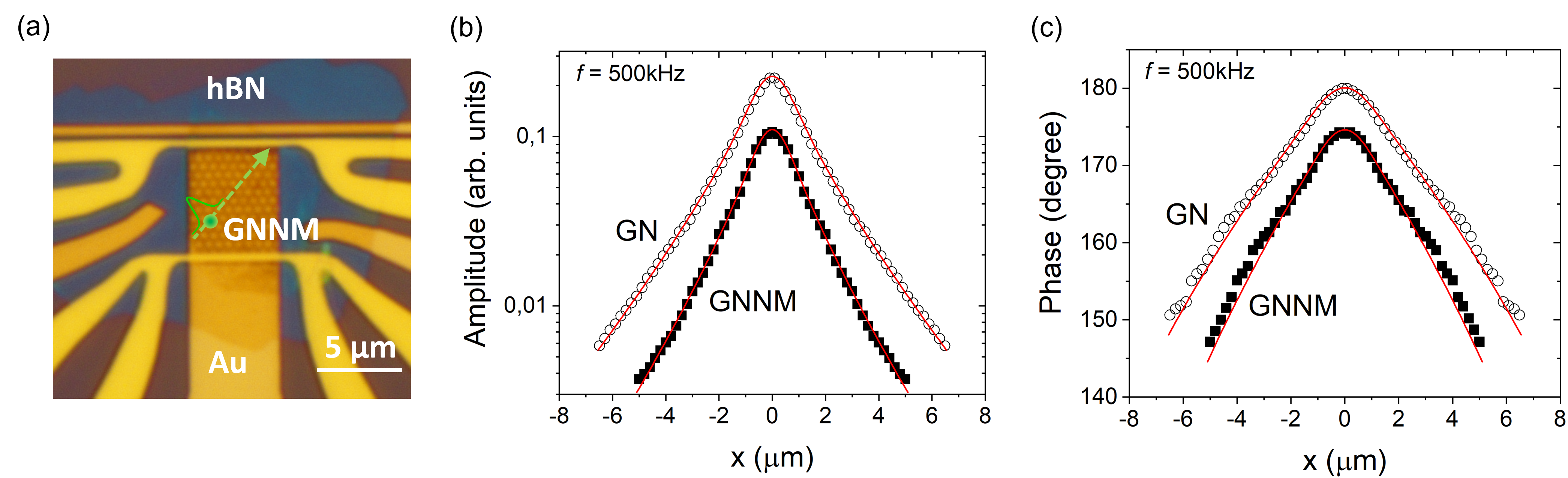}
\caption{\label{Fig4} (a) Optical image of the GNNM-based device for MTR measurements. The location and direction of the MTR scans is performed over GNNM is indicated by the green dotted
lines. The green spot indicates the zero scan position. The Gaussian beam profile of the green laser is schematically represented. (b) MTR amplitude and (c) phase data measured at 500~kHz for the GN (open symbols) and GNNM regions (solid symbols), along with the corresponding best-fit curves (solid lines). For clarity, the data and fits are vertically offset.}
\end{figure*}

Finally, we discuss the MTR measurements allowing the investigation of the thermal conductivity for two different representative samples, including a nanomeshed graphene (GNNM) and a pristine graphene (GN) flake. These samples were prepared simultaneously and share the same device structure as shown in Fig.~\ref{Fig1} (a), but with increased channel length, of $\sim 7~\mu$m, and comparable flake thicknesses, of approximately 3.5 nm and 4.3 nm for GNNM and GN, respectively. The longer channel facilitates more reliable laser scanning.
An optical microscope image of the GNNM-based sample used for the MTR investigation is presented in Fig.~\ref{Fig4} (a), where the laser scan region  measurement on the GNNM section is schematically indicated.
In MTR experiments, the measured AC electrical signal $S_{\mathrm{AC}}$ is proportional to the optical reflectivity variation $\Delta R_{\mathrm{probe}}$ of the probe beam induced by a temperature modulation at a fixed frequency $f$ in the heated area generated by the pump beam:
\begin{equation}
\label{eq:MTR}
S_{\mathrm{AC}} \sim \Delta R_{\mathrm{probe}} = \frac{\partial R_{\mathrm{probe}}}{\partial T} \Delta T
\end{equation}
Measuring $\Delta R_{\mathrm{probe}}$ allows to access the temperature profile induced by the pump beam.

We measure both the amplitude and phase of the probe signal as a function of the pump-probe spatial separation, using linear scans performed at different modulation frequencies ($100~\text{kHz} \leq f \leq 1~\text{MHz}$).
For each sample, the three different stacked materials (Au, hBN flake, and GN or GNNM flake) do not completely overlap. 
As a consequance, MTR measurements are performed at three different locations on the samples, where the local stack is composed of Au, Au/hBN, and Au/hBN/GN or GNNM at 100~kHz, 500~kHz and 1~MHz. These separate measurements enable the extraction of the thermal properties of each layer, accounting for the thermal contribution of the underlying layers. This approach clearly assumes that the thermal properties of a single flake are not spatially dependent.
Linear scans are symmetrically performed around the zero position (corresponding to pump-probe spot overlapping), and the scan range varies depending on the spatial constraints of each zone. When space is limited, scans are acquired in a half-range (e.g., $0 \leq x \leq 6~\mu$m or $-6~\mu\text{m} \leq x \leq 0$), and then mirrored about the origin. 
Heat transport is assumed to be diffusive and is described by the classical Fourier law. MTR measurements are analyzed by solving the heat diffusion equation in layered systems as outlined in Refs. \cite{fournier_Daniele_Marangolo_2020_measurement}. Note that the model does not account for the nanomesh geometry of the GNNM flake, which is instead treated as a uniform material. We consider this a reasonable approximation, since both the probe and pump beams have a diameter of $\sim 1.8~$µm. Given the nanomeshed geometry, the signal is therefore averaged over regions containing at least approximately 10 holes. Moreover, we observe reproducible results across different samples, as well as consistent signals when slightly shifting the scan position on the same sample, within the available space.
The fitting procedure determines the in-plane ($k_{\parallel}$) and out-of-plane ($k_{\perp}$) components of the thermal conductivity for each sample layer. These are the only free parameters. The best-fit values of $k_{\parallel}$ and $k_{\perp}$ are obtained by minimizing the differences between calculated and experimental amplitude and phase signals.

Figures~\ref{Fig4} (b) and (c) show as an example the experimental MTR amplitude and phase data at 500~kHz for the GN and GNNM flakes (open and solid symbols), respectively, along with the corresponding best-fit curves (solid lines). The data and fits are vertically offset for clarity. Additional measurements at 100~kHz and 1~MHz are reported in Fig.~S5 of the Supplementary Information, with thermal conductivity values extracted for each sample component summarized in Table~S1. 
The extracted values of the thermal conductivities for Au and hBN are consistent with literature \cite{rahimi_M_Sobnath_2023_complete, jiang_Puqing_Qian_2018_anisotropic, jaffe_Gabriel_R_2021_long, mason_SJ_Wesenberg_2020_violation}.
The average in-plane thermal conductivity values extracted from the best-fit across all frequencies are $k_{\parallel} \approx 2016$~ W/m K for GN and $k_{\parallel} \approx 723$~ W/m K for GNNM. We measure a very high thermal conductivity of the hBN-supported GN flake which is in good agreement with theoretical predictions, estimating values up to $\sim 1900$~ W/m K for graphene on hBN \cite{zhang_Zhongwei_Hu_2017_hexagonal, pak_Alexander_J_2016_theoretical, seol_Jae_Hun_2010_two}.
In particular, we clearly show that the nanomesh patterning of multilayer graphene reduces its in-plane thermal conductivity by approximately 65\%. The out-of-plane component $k_{\perp}$ remains nearly unchanged between the two samples, equal to $\approx 2$~ W/m K. Note, indeed, that $k_{\perp}$ includes thermal interface conductance contributions, which strongly depends on the fabrication process and are hard to control.
Note also that our MTR measurements are done at zero gate voltage, we expect our results to remain valid in the whole explored gate voltage range, corresponding to charge carrier densities ($n\sim 10^{12} cm^{-2}$) low enough to neglect a reduction of $k$ due to enhanced electron–phonon coupling \cite{tang_Jinyao_Wang_2010_holey}.
From the extracted $k$ values we evaluate a thermal conductivity ratio of $k_{\mathrm{GNNM}}/k_{\mathrm{GN}} \approx 0.36$. This value is lower than the expected reduction of a factor 0.55 predicted by the classical Maxwell-Eucken model. 
Being this model only valid when feature sizes exceed the phonon mean free path (MFP), observing a stronger reduction implies that the chosen neck width ($w\sim 160$~nm) is smaller than the effective phonon MFP, $l_{\mathrm{MFP}}$, in pristine multilayer GN. Hence, we can consider the used neck width as a lower limit of $l_{\mathrm{MFP}}$. Note that for single-layer GN supported by hBN, $l_{\mathrm{MFP}}$ is theoretically predicted up to approximately 700~nm \cite{zou_Ji_Hang_2017_phonon, Pak2016}. Further studies varying porosity are required to conclusively support this finding.
Our findings confirm that phonon transport in graphene thin flakes is more affected than electron transport by the applied nanomesh geometry, going behoynd the classical expectations. This provides evidence of an effective decoupling of the thermoelectric parameters through nanomeshing.

\section{\label{sec:five} Conclusions}
In conclusion, we have investigated the effect of nanostructuring, specifically nanomeshing via a hexagonal network of circular holes with a 160~nm neck width and a 360~nm diameter, on the electrical, thermoelectric, and thermal properties of multilayer graphene with thickness ranging from 4 to 6 nm. We have designed a device enabling the simultaneous investigation of electric and thermoelectric responses on the same graphene flake, where nanomeshing is applied to only a portion of it, thus allowing direct comparison. 
For the nanomeshed graphene, we have observed a reduced electrical conductivity in agreement with classical predictions for porous media and an enhanced Seebeck coefficient, resulting in an improvement of the maximum power factor $PF$ upon nanostructuring by more than $40 \%$.
Furthermore, thermal conductivity has been studied in analogous samples using the MTR technique. We have observed a reduction in the thermal conductivity in the graphene nanomeshed by a factor of almost 3 relative to the pristine graphene case, surpassing the classical Maxwell-Eucken limit and revealing the different lenght scales for electrons and phonons scattering.  Nanostructuring at low dimensions reveals as a powerful tool for engineering and control the thermoelectric properties of 2D materials, allowing an effective parameters  decoupling and opening the route to advanced energy conversion and thermal menagement solutions.

\begin{acknowledgments}
This research is funded by the
“Commissariat Général à l’Investissement d’Avenir” and
the French National Research Agency (ANR) (Grant No.
ANR-20-CE05-0045-01). This work is also partly supported
by the NANOFUTUR Equipex+ Program (ANR-
21-ESRE-0012) overseen by the French National Research
Agency as part of the “Programme d’Investissements
d’Avenir”.
Raman measurements were performed using a Witec alpha300R Raman-AFM platform supported by the IdEx Université Paris Cité, ANR-18-IDEX-0001 and by a SESAME grant from the region Île-de-France. The authors thank A. Alekin for technical support for the Raman measurements.
The authors thank P. Filloux, R. Bonnet et A. Saidoun for technical support in the clean room of the Laboratoire Matériaux et Phénomènes Quantiques (UMR 7162) at the Université Paris Cité.
\end{acknowledgments}

\subsection*{Conflict of interest}
The authors have no conflicts to disclose.

\section*{Data Availability}
The data that support the findings of this study are available from the corresponding authors upon reasonable request.


\end{document}